\documentclass[a4paper]{article}

\usepackage{INTERSPEECH2018}
\usepackage{multirow, hhline, stmaryrd, mathtools}
\DeclareMathOperator*{\argmax}{arg\,max}
\DeclareMathOperator*{\argmin}{arg\,min}

\title{Adversarial Feature-Mapping for Speech Enhancement}
\name{Zhong Meng$^{1,2}$ \thanks{Zhong Meng performed the work while he was a research intern
	at Microsoft AI and Research, Redmond, WA, USA.}, Jinyu Li$^1$, Yifan Gong$^1$, Biing-Hwang (Fred)
Juang$^2$}
\address{
  $^1$Microsoft AI and Research, Redmond, WA, USA\\
  $^2$Georgia Institute of Technology, Atlanta, GA, USA}
  \email{zhongmeng@gatech.edu, \{jinyli, yifan.gong\}@microsoft.com, juang@ece.gatech.edu}

\begin{document}

\maketitle
\begin{abstract}

Feature-mapping with deep neural networks is commonly used for single-channel speech enhancement, in which a feature-mapping network directly transforms the noisy features to the corresponding enhanced ones and is trained to minimize the mean square errors between the enhanced and clean features. In this paper, we propose an adversarial feature-mapping (AFM) method for speech enhancement which advances the feature-mapping approach with adversarial learning. An additional discriminator network is introduced to distinguish the enhanced features from the real clean ones. The two networks are jointly optimized to minimize the feature-mapping loss and simultaneously mini-maximize the discrimination loss. The distribution of the enhanced features is further pushed towards that of the clean features through this adversarial multi-task training. To achieve better performance on ASR task, senone-aware (SA) AFM is further proposed in which an acoustic model network is jointly trained with the feature-mapping and discriminator networks to optimize the senone classification loss in addition to the AFM losses. Evaluated on the CHiME-3 dataset, the proposed AFM achieves 16.95\% and 5.27\% relative word error rate (WER) improvements over the real noisy data and the feature-mapping baseline respectively and the SA-AFM achieves 9.85\% relative WER improvement over the multi-conditional acoustic model.

\end{abstract}

\section{Introduction}

Single-channel speech enhancement aims at attenuating the noise component
of noisy speech to increase the intelligibility and perceived quality of
the speech component \cite{loizou2013speech}.  It is commonly used to
improve the quality of mobile speech communication in noisy environments
and enhance the speech signal before amplification in hearing aids and
cochlear implants. 
More importantly, speech enhancement is widely applied as a front-end
pre-processing stage to improve the performance of automatic speech
recognition (ASR) \cite{hinton2012deep, jaitly2012application,
sainath2011making, deng2013recent, yu2017recent} and speaker recognition
under noisy conditions \cite{Li14overview, Li15robust}.

With the advance of deep learning, deep neural network (DNN) based
approaches have achieved great success in single-channel speech enhancement.
The mask learning approach \cite{narayanan2013ideal, wang2014training,
weninger2015speech} is proposed to estimate the ideal ratio mask or ideal
binary mask based on noisy input features using a DNN. The mask is used to
filter out the noise from the noisy speech and recover the clean speech.
However, it has the presumption that the scale of the masked signal is the
same as the clean target and the noise is strictly additive and removable
by the masking procedure which is generally not true for real recorded
stereo data.  To deal with this problem, the feature-mapping approach
\cite{xu2015regression, lu2013speech, maas2012recurrent, feng2014speech,
weninger2014single, chen2017improving} is proposed to train a feature-mapping network that directly
transforms the noisy features to enhanced ones. The feature-mapping network serves
as a non-linear regression function trained to minimize the feature-mapping
loss, i.e., the mean square error (MSE) between the enhanced features and
the paralleled clean ones. The application of MSE estimator is based on the
homoscedasticity and no auto-correlation assumption of the noise, i.e., the
noise needs to have the same variance for each noisy feature and the noise needs to be uncorrelated between different noisy features \cite{freedman2009statistical}. This assumption is
in general violated for real speech signal (a kind of time series data) under
non-stationary unknown noise.



Recently, adversarial training \cite{gan} has become a hot topic in deep
learning with its great success in estimating generative models. It was first
applied to image generation \cite{radford2015unsupervised, denton2015deep},
image-to-image translation \cite{isola2017imagetoimage, zhu2017unpaired}
and representation learning \cite{chen2016infogan}.  In speech area, it has
been applied to speech enhancement \cite{pascual2017segan,
donahue2017exploring, mimura2017cross, meng2018cycle}, voice conversion
\cite{kaneko2017parallel, hsu2017voice}, acoustic model adaptation
\cite{sun2017unsupervised, meng2017unsupervised, meng2018adversarial}, noise-robust
\cite{grl_shinohara, grl_serdyuk} and speaker-invariant
\cite{saon2017english, meng2018speaker} ASR using gradient reversal layer (GRL)
\cite{ganin2015unsupervised}. In these works, adversarial training is used to learn
a feature or an intermediate representation in DNN that is invariant to
the shift among different domains (e.g., environments, speakers, image styles, etc.). In other words, a generator network is trained to map
data from different domains to the features with similar distributions via
adversarial learning. 

Inspired by this, we advance the feature-mapping
approach with adversarial learning to further diminish the discrepancy between the
distributions of the clean features and the enhanced features generated by the
feature-mapping network given non-stationary and auto-correlated noise at the input. We call this method adversarial feature-mapping (AFM) for speech
enhancement. In AFM, an additional discriminator network is introduced to
distinguish the enhanced features from the real clean ones.  The feature-mapping
network and the discriminator network are jointly trained to minimize the
feature-mapping loss and simultaneously mini-maximize the discrimination
loss with adversarial multi-task learning. With AFM, the
feature-mapping network can generate pseudo-clean features that the
discriminator can hardly tell whether they are real clean features or not. To achieve better performance on ASR task, senone-aware adversarial feature-mapping (SA-AFM) is proposed in which an acoustic model network is introduced and is jointly trained with the feature-mapping and discriminator networks to optimize the senone classification loss in addition to the feature-mapping and discrimination losses.


Note that AFM is different from
\cite{donahue2017exploring} in that: (1) In AFM, the inputs to the
discriminator are enhanced and clean features while in
\cite{donahue2017exploring} the inputs to the discriminator are the
concatenation of enhanced and noisy features and the concatenation of clean
and noisy features. (2) The primary task of AFM is feature-mapping, i.e., to
minimize the $L_2$ distance (MSE) between enhanced and clean features and it is advanced
with adversarial learning to further reduce the discrepancy between the distributions of the enhanced and clean features while in
\cite{donahue2017exploring} the primary task is to generate enhanced
features that are similar to clean features with generative adversarial
network (GAN) and it is regularized with the minimization of $L_1$ distance
between noisy and enhanced features. (3) AFM performs adversarial multi-task training using GRL
method as in \cite{ganin2015unsupervised} while \cite{donahue2017exploring}
conducts conditional GAN iterative optimization as in \cite{gan}. (4) In
this paper, AFM uses long short-term memory (LSTM)-recurrent neural
networks (RNNs) \cite{sak2014long, meng2017deep, erdogan2016multi} to generate the enhanced features and a feed-forward DNN as
the discriminator while \cite{donahue2017exploring} uses convolutional
neural networks for both.

We perform ASR experiments with features enhanced by AFM on
CHiME-3 dataset \cite{barker2015third}. Evaluated on a clean acoustic
model, AFM achieves 16.95\% and 5.27\% relative word error rate (WER) improvements
respectively over the noisy features and feature-mapping baseline and the SA-AFM achieves 9.85\% relative WER improvement over the multi-conditional acoustic model.



\section{Adversarial Feature-Mapping Speech Enhancement}
\label{sec:cse}

With feature-mapping approach for speech enhancement, we are given a
sequence of noisy speech features $X=\{x_1, \ldots, x_T\}$ and a sequence
of clean speech features $Y=\{y_1, \ldots, y_T\}$ as the training data. $X$
and $Y$ are \emph{parallel} to each other, i.e., each pair of $x_i$ and
$y_i$ is frame-by-frame synchronized. The goal of speech enhancement
is to learn a non-linear feature-mapping network $F$ with parameters $\theta_f$ that transforms 
$X$ to a sequence of enhanced features $\hat{Y}=\{\hat{y}_1,
\ldots, \hat{y}_T\}$ such that the distribution of $\hat{Y}$ is as close to that of $Y$ as possible:
\begin{align}
	& \hat{y_i} = F(x_i), \quad i = 1, \ldots, T \\
	& P_{\hat{Y}}(\hat{y}) \rightarrow P_Y(y).
	\label{eqn:fm}
\end{align}
To achieve that, we minimize the noisy-to-clean feature-mapping loss
$\mathcal{L}_{F}(\theta_f)$,
which is commonly defined as the MSE between 
$\hat{Y}$ and $Y$ as follows:
\begin{align}
	\mathcal{L}_{F}(\theta_f) &= 
	\frac{1}{T}\sum_{i=1}^T (\hat{y}_i -y_i)^2 =
	\frac{1}{T}\sum_{i=1}^T \left[F(x_i) - y_i\right]^2.
	\label{eqn:loss_f}
\end{align}

However, the MSE that feature-mapping approach minimizes is based on the
homoscedasticity and no auto-correlation assumption of the noise, i.e., the
noise has the same variance for each noisy feature and the noise is
uncorrelated between different noisy features. This assumption is
in general invalid for real speech signal (time series data) under
non-stationary unknown noise.  To address this problem, we further advance the feature-mapping network with an additional discriminator network and perform
adversarial multi-task training to further reduce the discrepancy between the
distribution of enhanced features and the clean ones given non-stationary
and auto-correlated noise is at the input.

\begin{figure}[htpb!]
    \centering
    \includegraphics[width=0.65\columnwidth]{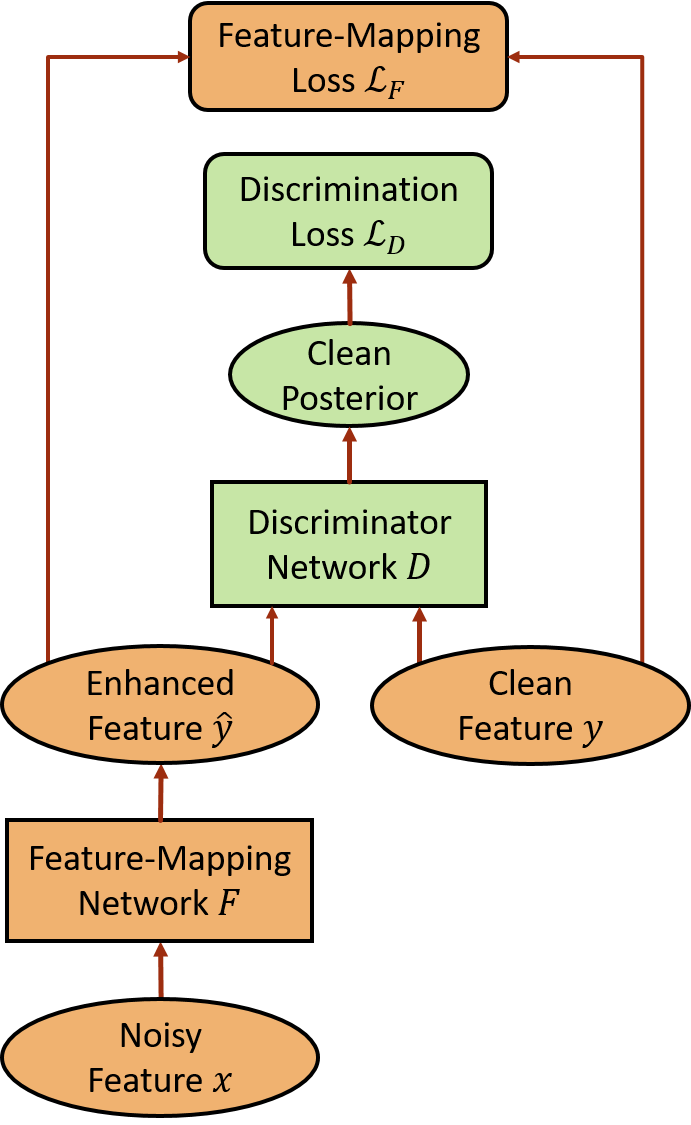}
     \vspace{-0.0cm}
    \caption{\small The framework of AFM for speech enhancement.
	    }
    \label{fig:afm}
\end{figure}

As shown in Fig. \ref{fig:afm}, the discriminator network $D$ with
parameters $\theta_d$ takes
enhanced features $\hat{Y}$ and clean features $Y$ as the input and outputs the posterior
probability that an input feature belongs to the clean set, i.e.,
\begin{align}
	P(y_i \in \mathbb{C})& = D(y_i)  \\
	P(\hat{y}_i \in \mathbb{E})& = 1 - D(\hat{y}_i)
	\label{eqn:discriminator}
\end{align}
where $\mathbb{C}$ and $\mathbb{E}$ denote the
sets of clean and enhanced features respectively.  The
discrimination losses $\mathcal{L}_{D}(\theta_f, \theta_d)$ for the $D$ is formulated below using cross-entropy: 
\begin{align}
	& \mathcal{L}_{D}(\theta_f, \theta_d) = \frac{1}{T}\sum_{i = 1}^{T} \left[ \log
	P(y_i \in \mathbb{C}) + \log P(\hat{y}_i \in \mathbb{E}) \right] \nonumber \\
        & = \frac{1}{T}\sum_{i = 1}^{T} \log D(y_i) + \log
	\left[1 - D(F(x_i))\right]. \label{eqn:loss_d}
\end{align}

To make the distribution of the enhanced features $\hat{Y}$ similar to that
of the clean ones $Y$, we perform adversarial training of $F$ and $D$, i.e, we
minimize $\mathcal{L}_{D}(\theta_f, \theta_d)$ with respect to $\theta_d$
and maximize $\mathcal{L}_{D}(\theta_f, \theta_d)$ with respect to $\theta_f$.
This minimax competition will first increase the generation capability of $F$ and the discrimination capability of $D$ and will eventually converge to the point where the $F$ generates
extremely confusing enhanced features that $D$ is unable to distinguish
whether it is a clean feature or not.

The total loss of AFM $\mathcal{L}_{AFM}(\theta_f, \theta_d)$ is
formulated as the weighted sum of the
feature-mapping loss and the discrimination loss below:
\begin{align}
	&\mathcal{L}_{\text{AFM}}(\theta_f, \theta_d) =
	\mathcal{L}_{F}(\theta_f)
	- \lambda \mathcal{L}_{D}(\theta_f, \theta_d)
	\label{eqn:loss_afm}
\end{align}
where $\lambda > 0$ is the gradient reversal coefficient that controls the trade-off between the feature-mapping loss and the
discrimination loss in Eq. \eqref{eqn:loss_f} and Eq. \eqref{eqn:loss_d} respectively.

$F$ and $D$ are jointly trained to optimize the total loss through adversarial
multi-task learning as follows:
\begin{align}
    \hat{\theta}_f = \argmin_{\theta_f}
    \mathcal{L}_{\text{AFM}}(\theta_f, \hat{\theta}_d) \label{eqn:min_f} \\
    \hat{\theta}_d = \argmax_{\theta_d}
    \mathcal{L}_{\text{AFM}}(\hat{\theta}_f, \theta_d)
    \label{eqn:max_d}
\end{align}
where $\hat{\theta}_f$ and $\hat{\theta}_d$ are optimal parameters for
$F$ and $D$ respectively and are updated as follows via back propagation through
time (BPTT) with stochastic gradient descent (SGD):
\begin{align}
	& \theta_f \leftarrow \theta_f - \mu \left[ \frac{\partial
		\mathcal{L}_{\text{F}}(\theta_f)}{\partial \theta_f} - \lambda \frac{\partial
			\mathcal{L}_{\text{D}}(\theta_f, \theta_d)}{\partial
			\theta_f}
		\right]
		\label{eqn:grad_f} \\
	& \theta_d \leftarrow \theta_d - \mu \frac{\partial
		\mathcal{L}_{\text{D}}(\theta_f, \theta_d)}{\partial \theta_d}
		\label{eqn:grad_d}
\end{align}
where $\mu$ is the learning rate.

Note that the negative coefficient $-\lambda$ in Eq. \eqref{eqn:grad_f}
induces reversed gradient that maximizes
$\mathcal{L}_{D}(\theta_f, \theta_d)$ in Eq. \eqref{eqn:loss_d}
and makes the enhanced features similar to the real clean ones. 
Without the reversal gradient, SGD would make the
enhanced features different from the clean ones in order to minimize Eq.
\eqref{eqn:loss_d}. For easy implementation, gradient reversal layer is introduced in
\cite{ganin2015unsupervised}, which acts as an identity transform in the forward propagation
and multiplies the gradient by $-\lambda$ during the backward propagation.
During testing, only the optimized feature-mapping network $F$ is used to
generate the enhanced features given the noisy test features.


\section{Senone-Aware Adversarial Feature-Mapping Enhancement}
\label{sec:saafm}
For AFM speech enhancement, we only need parallel clean and noisy speech for training and we do not need any information about the content of the speech, i.e., the transcription. With the goal of improving the intelligibility and perceived quality of the speech, AFM can be widely used in a broad range of applications including ASR, mobile communication, hearing aids, cochlear implants, etc. However, for the most important ASR task, AFM does not necessarily lead to the best WER performance because its feature-mapping and discrimination objectives are not directly related to the speech units (i.e., word, phoneme, senone, etc.) classification. In fact, with AFM, some decision boundaries among speech units may be distorted in searching for an optimal separation between speech and noise. 

To compensate for this mismatch, we incorporate a DNN acoustic model into the AFM framework and propose the senone-aware adversarial feature-mapping (SA-AFM), in which the acoustic model network $M$, feature-mapping network $F$ and the discriminator network $D$ are trained to jointly optimize the primary task of feature-mapping, secondary task of the third task of clean/enhanced data discrimination and the third task of senone classification in an adversarial fashion. The transcription of the parallel clean and noisy training utterances is required for SA-AFM speech enhancement.

\begin{figure}[htpb!]
    \centering
    \includegraphics[width=0.8\columnwidth]{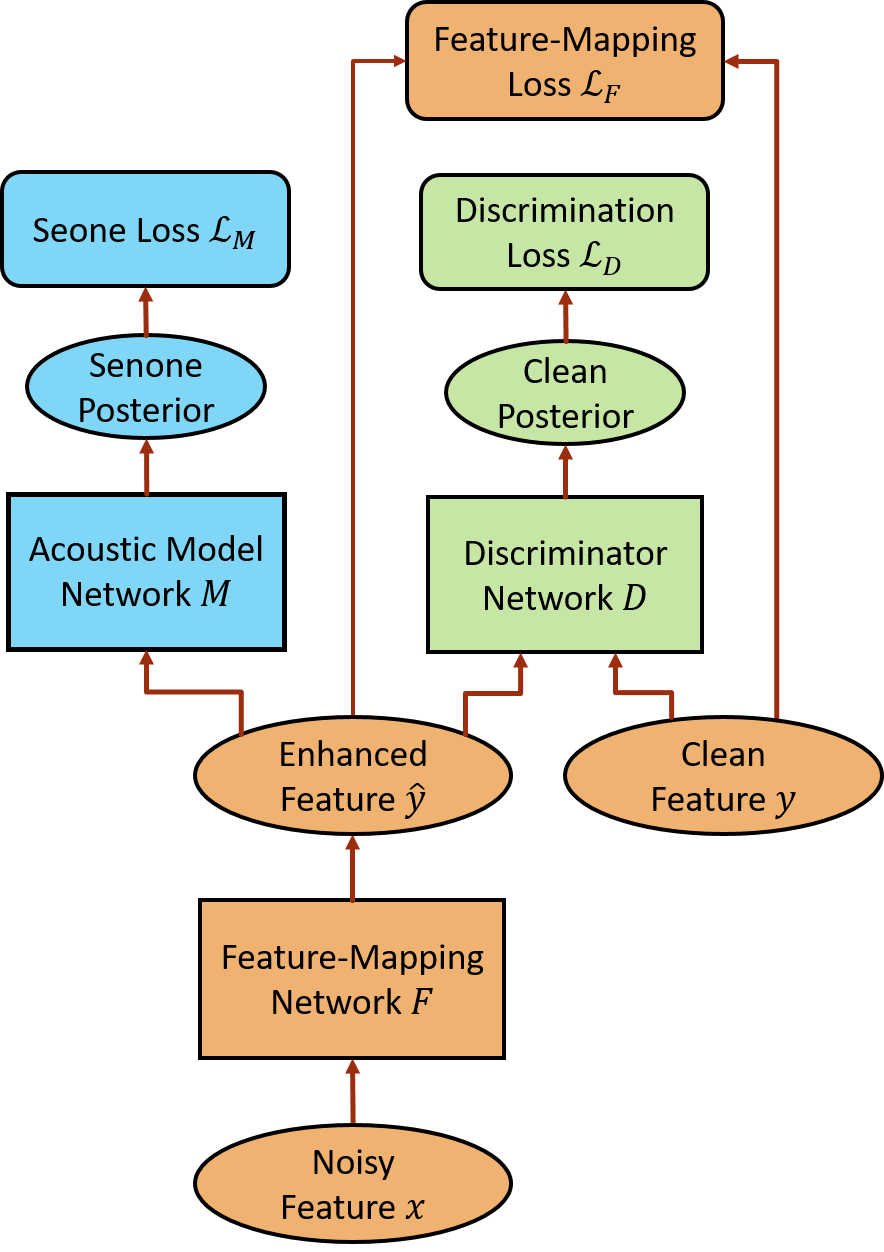}
     \vspace{-0.0cm}
    \caption{\small The framework of SA-AFM for speech enhancement.
	    }
    \label{fig:saafm}
    \vspace{-20pt}
\end{figure}

Specifically, as shown in Fig. \ref{fig:saafm}, the acoustic model network $M$ with parameters $\theta_m$ takes in the enhanced features $\hat{Y}$ as the input and predicts the senone posteriors $P(q|\hat{y}_i; \theta_y), q\in \mathcal{Q}$ as follows:
\begin{align}
	M(\hat{y}_i) =  P(q | \hat{y}_i; \theta_m)，
	\label{eqn:senone_classify_1}
\end{align}
after the integration with feature-mapping network $F$, we have
\begin{align}
	M(F(x_i)) = P(q | x_i; \theta_f, \theta_m).
	\label{eqn:senone_classify_2}
\end{align}

We want to make the enhanced features $\hat{Y}$ senone-discriminative by
minimizing the cross-entropy loss between the predicted senone posteriors
and the senone labels as follows:
\begin{align}
	\mathcal{L}_{\text{M}}(\theta_m, \theta_f) & = - \frac{1}{T}\sum_{i = 1}^{T} \log P(s_i |
	x_i;\theta_f, \theta_m) \nonumber \\
	& = -\frac{1}{T}\sum_{i = 1}^{T} \log M(F(x_i))
	\label{eqn:loss_m}
\end{align}
where $S$ is a sequence of senone labels $S=\{s_{1},\ldots, s_{T}\}$
aligned with the noisy data $X$ and enhanced data $\hat{Y}$. 

Simultaneously, we minimize feature-mapping loss $\mathcal{L}_{F}(\theta_f)$ defined in Eq. \eqref{eqn:loss_f} with respect to $F$ and perform adversarial training of $F$ and $D$, i.e, we
minimize $\mathcal{L}_{D}(\theta_f, \theta_d)$ defined in Eq. \eqref{eqn:loss_d} with respect to $\theta_d$
and maximize $\mathcal{L}_{D}(\theta_f, \theta_d)$ with respect to $\theta_f$, to make the distribution of the enhanced features $\hat{Y}$ similar to that
of the clean ones $Y$.
 
The total loss of SA-AFM $\mathcal{L}_{\text{SA-AFM}}(\theta_m, \theta_f, \theta_d)$ is
formulated as the weighted sum of $\mathcal{L}_{F}(\theta_f)$, $\mathcal{L}_{D}(\theta_f, \theta_d)$ and the senone classification loss $\mathcal{L}_{\text{M}}(\theta_m, \theta_f)$ as follows:
\begin{align}
	\mathcal{L}_{\text{SA-AFM}}(\theta_f, \theta_d, \theta_m) = &
    \mathcal{L}_{F}(\theta_f)- \lambda_1 \mathcal{L}_{D}(\theta_f, \theta_d) \nonumber \\
	& + \lambda_2 \mathcal{L}_{M}(\theta_f, \theta_m)
	\label{eqn:loss_saafm}
\end{align}
where $\lambda_1 > 0$ is the gradient reversal coefficient that controls the trade-off between $\mathcal{L}_{F}(\theta_f)$ and $\mathcal{L}_{D}(\theta_f, \theta_d)$, and $\lambda_2 > 0$ is the weight for $\mathcal{L}_{M}(\theta_f, \theta_m)$. 

$F$, $D$ and $M$ are jointly trained to optimize the total loss through adversarial
multi-task learning as follows:
\begin{align}
    (\hat{\theta}_f, \hat{\theta}_m) = \argmin_{\theta_f, \theta_m}
    \mathcal{L}_{\text{SA-AFM}}(\theta_f, \hat{\theta}_d, \theta_m) \label{eqn:saafm_fm} \\
    \hat{\theta}_d = \argmax_{\theta_d}
    \mathcal{L}_{\text{SA-AFM}}(\hat{\theta}_f, \theta_d, \hat{\theta}_m)
    \label{eqn:saafm_d}
\end{align}
where $\hat{\theta}_f$, $\hat{\theta}_d$ and $\hat{\theta}_m$ are optimal parameters for
$F$, $D$ and $M$ respectively and are updated as follows via BPTT with SGD as in Eq. \eqref{eqn:grad_f}, Eq. \eqref{eqn:grad_d} and Eq. \eqref{eqn:grad_m} below:
\begin{align}
	\theta_m \leftarrow \theta_m - \mu \frac{\partial
		\mathcal{L}_{\text{M}}(\theta_f, \theta_m)}{\partial \theta_m}.
		\label{eqn:grad_m}
\end{align}
During decoding, only the optimized feature-mapping network $F$ and acoustic model network $M$ are used to take in the noisy test features and generate the acoustic scores.

\section{Experiments}

In the experiments, we train the feature-mapping network $F$ with the
parallel clean and noisy training utterances in CHiME-3 dataset \cite{barker2015third} using different
methods.  The real far-field noisy speech from the 5th microphone channel
in CHiME-3 development data set is used for testing. We use a pre-trained
clean DNN acoustic model to evaluate the ASR WER performance of the test
features enhanced by $F$. The standard WSJ 3-gram language model with 5K-word
lexicon is used in our experiments.



\subsection{Feedforward DNN Acoustic Model}
\label{sec:dnn_am}

To evaluate the ASR performance of the features enhanced by AFM,
we first train a feedforward DNN-hidden Markov model (HMM) acoustic model using 8738
clean training utterances in CHiME-3 with cross-entropy criterion. The
29-dimensional log Mel filterbank (LFB)  features together with 1st and 2nd order
delta features (totally 87-dimensional) are extracted. Each feature frame
is spliced together with 5 left and 5 right context frames to form a
957-dimensional feature. The spliced features are fed as the input of the
feed-forward DNN after global mean and variance normalization. The DNN has
7 hidden layers with 2048 hidden units for each layer. The output layer of
the DNN has 3012 output units corresponding to 3012 senone labels.
Senone-level forced alignment of the clean data is generated using a
Gaussian mixture model-HMM system. A WER of 29.44\% is achieved when
evaluating the clean DNN acoustic model on the test data.

\subsection{Adversarial Feature-Mapping Speech Enhancement}
\label{sec:exp_afm}

We use parallel data consisting of 8738 pairs of noisy and clean utterances
in CHiME-3 as the training data.  The 29-dimensional LFB features are extracted
for the training data. For the noisy data, the 29-dimensional LFB
features are appended with 1st and 2nd order delta features to form
87-dimensional feature vectors. $F$ is an LSTM-RNN with 2 hidden layers and
512 units for each hidden layer. A 256-dimensional projection layer is
inserted on top of each hidden layer to reduce the number of parameters.
$F$ has 87 input units and 29 output units. The features are globally mean
and variance normalized before fed into $F$. The discriminator $D$ is a
feedforward DNN with 2 hidden layers and 512 units in each hidden layer.
$D$ has 29 input units and one output unit.

We first train $F$ with 87-dimensional LFB features as the input and
29-dimensional LFB features as the target to minimize the feature-mapping
loss $\mathcal{L}(\theta_f)$ in Eq. \eqref{eqn:loss_f}. This serves as the
feature-mapping baseline. Evaluated on clean
DNN acoustic model trained in Section \ref{sec:dnn_am}, the feature-mapping
enhanced features achieve 25.81\% WER which is 12.33\% relative improvement
over the noisy features. Then we jointly
train $F$ and $D$ to optimize $\mathcal{L}_{\text{AFM}}(\theta_f,
\theta_d)$ as in Eq. \eqref{eqn:loss_afm} using the same input features
and targets. The gradient reversal coefficient $\lambda$ is fixed at $60$ and the learning rate is $5\times 10^{-7}$ with a momentum of $0.5$ in the experiments. As shown in Table \ref{table:wer_afm}, AFM enhanced features
achieve 24.45\% WER which is 16.95 \% and 5.27\% relative improvements over the 
noisy features and feature-mapping baseline.

\begin{table}[h]
\centering
\begin{tabular}[c]{c|c|c|c|c|c}
	\hline
	\hline
	Test Data & BUS & CAF & PED & STR & Avg.\\
	\hline
	Noisy & 36.25 & 31.78 & 22.76 & 27.18 & 29.44\\
	\hline
	FM & 31.35 & 28.64 & 19.80 & 23.61 & 25.81
	\\
	\hline
        AFM & 30.97 & 26.09 & 18.40 & 22.53 & 24.45 \\
	\hline
	\hline
\end{tabular}
  \caption{The ASR WER (\%) performance of real noisy dev set in CHiME-3
	  enhanced by different methods evaluated
  on a clean DNN acoustic model. FM represents feature-mapping.}
\label{table:wer_afm}
\vspace{-20pt}
\end{table}

\subsection{Senone-Aware Adversarial Feature-Mapping Speech Enhancement}
The SA-AFM experiment is conducted on top of the AFM system described in Section \ref{sec:exp_afm}. In addition to the LSTM $F$ and feedforward DNN $D$, we train a multi-conditional LSTM acoustic model $M$ using both the 8738 clean and 8738 noisy training utterances in CHiME-3 dataset. The LSTM $M$ has 4 hidden layers with 1024 units in each layer. A 512-dimensional projection layer is inserted on top each hidden layer to reduce the number of parameters. The output layer has 3012 output units predicting senone posteriors. The senone-level forced alignment of the training data is generated using a GMM-HMM system. As shown in Table \ref{table:wer_saafm}, the multi-conditional acoustic model achieves 19.28\% WER on CHiME-3 simulated dev set.

\begin{table}[h]
\centering
\begin{tabular}[c]{c|c|c|c|c|c}
	\hline
	\hline
	System & BUS & CAF & PED & STR & Avg. \\
	\hline
	Multi-Condition & 18.44 & 23.37 & 16.81 & 18.50 & 19.28 \\
	\hline
	SA-FM & 18.19 & 22.29 & 15.31 & 18.26 & 18.51 \\
	\hline
    SA-AFM & 17.02 & 21.01 & 14.41 & 17.13 & 17.38 \\
	\hline
	\hline
\end{tabular}
  \caption{The ASR WER (\%) performance of simulated noisy dev set in CHiME-3
	  by using multi-conditional acoustic model and different enhancement methods.}
\label{table:wer_saafm}
\vspace{-10pt}
\end{table}

Then we perform senone-aware feature-mapping (SA-FM) by jointly training $F$ and $M$ to optimize the feature-mapping loss and the senone classification loss in which $M$ takes the enhanced LFB features generated by $F$ as the input to predict the senone posteriors. The SA-FM achieves 18.51\% WER on the same testing data. Finally, SA-AFM is performed as described in Section \ref{sec:saafm} and it achieves 17.38\% WER which is 9.85\% and 6.10\% relative improvements over the multi-conditional acoustic model and SA-FM baseline.

\section{Conclusions}
In this paper, we advance feature-mapping approach with adversarial learning by
proposing AFM method for speech enhancement. In AFM, we have a
feature-mapping network $F$ that transforms the noisy speech features to
clean ones with parallel noisy and clean training data and a
discriminator $D$ that distinguishes the enhanced features from the clean
ones. $F$ and $D$ are jointly trained to minimize the feature-mapping loss
(i.e., MSE) and simultaneously mini-maximize the discrimination loss. On top of
feature-mapping, AFM pushes the distribution of the enhanced
features further towards that of the clean features with adversarial
multi-task learning.To achieve better performance on ASR task, SA-AFM is 
further proposed  to optimize the senone classification loss
in addition to the AFM losses.

We perform ASR experiments with features enhanced by AFM on CHiME-3
dataset. AFM achieves 16.95\% and 5.27\% relative WER improvements over the
noisy features and feature-mapping baseline when evaluated on a clean DNN
acoustic model. Furthermore, the proposed SA-AFM achieves 9.85\% relative WER improvement over the multi-conditional acoustic model. As we show in \cite{ts_adapt}, teacher-student (T/S) learning \cite{ts_learning} is better for robust model adaptation without the need of transcription. We are now working on the combination of AFM with T/S learning to further improve the ASR model performance.

\clearpage

\bibliographystyle{IEEEtran}

\bibliography{mybib}


\end{document}